# Evaluation of techniques for predicting seizure Build up

Amira Hajjeji[1] and Nawel Jmail[2,3] and Abir Hadriche[1,3,4*] and Amal Ncibi[1] and Chokri Ben Amar[4]

[1]Université de Gabes, ISIMG, Tunisie
[2]Université de Sfax, MIRACL. LAb, Sfax, Tunisie.
[3]Université de Sfax, centre de recherche numérique de Sfax, Tunisie.
[4]Université de Sfax, ENIS, REGIM. LAb, Route de Soukra Km 3.5, BP. 1173 – 3038, Sfax, Tunisie.
[*]abir.hadriche.tn@ieee.org

**Abstract.** The analysis of electrophysiological signal of scalp: EEG (electroencephalography), MEG (magnetoencephalography) and depth (intracerebral EEG) IEEG is a way to delimit "epileptogenic zone" (EZ). These epileptic signals present two different activities (oscillations and spikes) which can be overlapped in the time frequency plane. Automatic recognition of epileptic seizure occurrence needs several preprocessing steps. In this study, we evaluated two filtering techniques: the stationary wavelet transforms (SWT) and the Despikifying in order to extract pre ictal gamma oscillations (bio markers of seizure build up). Then, we used a temporal basis set of Jmail et al 2017 as a preprocessing step to evaluate the performance of both technique. Moreover, we used time-frequency and spatio-temporal mapping of simulated and real data for both techniques in order to predict seizure build up (in time and space). We concluded that SWT can detect the oscillations, but Despikyfying is more robust than SWT in reconstructing pure pre ictal gamma oscillations and hence in predicting seizure build up.

## 1 Introduction

Despite development in neurological disease diagnosis, the analysis of electrophysiological signal remains one of the important techniques. Magnetic fields MEG (magnetoencephalography) and electrical potentials EEG (electroencephalography) are registered in a non-invasive way, moreover intracerebral EEG (IEEG) is recorded as a traumatic technique straightly from the brain. In epilepsy, these signals depict spikes, oscillations at different frequencies and overlapped spikes and oscillations (1) (2). Excessive discharges could be registered into three stages (3) (4):'ictal' (during seizure) 'pre ictal' (before seizure) and 'inter ictal', (between seizure) state through analyzing both standard biomarkers: epileptic spikes (sharp transient) and oscillations either in the gamma band or in high frequency oscillations HFO (5). For better epilepsy diagnosis and epileptic patient well being, pre ictal state (where activities lead to a seizure) deserves a careful investigation (6) (7) (8).

In fact, studying IEEG pre ictal signals proves the presence of different activities (oscillatory and transient) either separated and /or overlapped in the time-frequency domain. The major problem here is to choose the accurate filtering technique in order to separate between these activities (oscillations versus spikes) in order to define their responsible cortical generators (9).

Several filtering techniques are used and evaluated in separation between oscillatory and spiky events such as: Matching Pursuit (MP), Filter Response Finite Impulse (FIR), Stationary Wavelet Transform (SWT) and Despikifying (1) (2) (10) (11) (12) (13) (14) (15).

Our goal in this study is to detect non contaminated pre ictal gamma oscillations by spiky events in order to predict seizure build up through a time and space mapping technique. To ensure a good separation between spikes and pre ictal gamma oscillations, advanced filtering techniques have been evaluated which are the stationary wavelet transform "SWT" and Despikifying.

## 2   Materials and Methods

Firstly, we will describe our database, and then we will apply our preprocessing chain proposed in (11) to evaluate robustness of our techniques in terms of pure pre ictal gamma oscillations reconstruction accuracy and hence recognition of seizure build up . Finally, we will present results and conclusion proving that SWT could separate pre ictal gamma oscillations from spiky events, but with a remained spiky part and Despikifying is more robust than SWT in reconstructing pure gamma waves non contaminated by spikes events.

### 2.1   Materials

All our exploited data and preprocessing steps are analysed on "MATLAB" Mathworks, Natick, MA, and EEGlab toolbox.

**Simulated data:** Our simulated data was sampled at 1000 Hz, composed of 6 channels depicting both states pre ictal and ictal states. Each channel shows oscillatory and spiky events separated and overlapped with a white noise as in (11) where signal to noise ratio was set to 5 dB.

**Real data:** Our analysed intra-cerebral EEG signal was a pre surgical acquisition of a pharmaco-resistant epileptic patient that depicts pre ictal gamma oscillations and spiky events. Our subject was diagnosed with a right occipito-temporal junction a symptomatic focal cortical dysplasia. All the pre-treatment and acquisition phases were done under supervision of an expert neurologist at the department of neurology of La Timone hospital in Marseille (11).

Our IEEG data was recorded on a Deltamed system with a sampling rate set to 128 Hz, composed of 96 channels and 4 events each one lasting 3 min and 20 second.

## 2.2 Methods

**Stationary Wavelet Transform:** is obtained from the Discrete Wavelet Transform "DWT" by a step of convolution then a decimation. SWT has been evaluated and used in various applications such as filtering and automatic detection. SWT has been proven very useful in analysis of electrophysiological signals. In fact, the best advantage of this filtering method is its property of time invariance (16) and its ability in overcoming frequency overlapping between physiological biomarkers (2).

SWT decomposes original signal *s(t)* at each step *k* and scale *j*, then projects it on a scale function (Mother wavelet function) dilated, compressed and translated according to two equations (1) and (2):

$$C_{j,k} = \langle s(t), \varphi_{j,k(t)} \rangle, \quad \varphi_{j,k(t)} = 2^{-j}\varphi(2^{-j}(t-k)) \tag{1}$$

$$W_{j,k} = \langle s(t), \psi_k(t) \rangle, \quad \psi_k(t) = 2^{-j}\psi(2^{-j}(t-k)) \tag{2}$$

With $C_{j,k}$ are approximation coefficients and $W_{j,k}$ are detail coefficients at the $j^{th}$ decomposition level and $\varphi_{j,k(t)}$ is the scale function (it's a low pass filter).

In this study, the separation between spiky events and pre ictal gamma oscillations was occurred using "Symlets" wavelets with 6 level of decomposition. We used *swt* Matlab function for decomposition step and *iswt* Matlab function for reconstruction of pure gamma oscillations after a thresholding step on the basis of rectangular masks as in (2).

**Despikifying:** is a nonlinear filtering technique that allows the removal of spiky events from original signal. In fact despikifying was tested and validated in (17) (11) as a filtering technique that delineates spiky events even overlapped with oscillatory activities. Despikifying is based on modeling spiky shapes using a Gaussian function version fitted non-linearly to the data, then projecting our original signal on the obtained spiky model and finally subtracting spikes from original signal in order to get only pure pre ictal gamma oscillations. Spiky model was fitted per each channel according to this equation:

$$G(t, a, b, \gamma) = \begin{cases} -Ae^{\frac{[(t-a)*\gamma]^2}{b}}, & t<a \\ Ae^{\frac{[(t-a)/\gamma]^2}{b}}, & t>a \end{cases} \tag{3}$$

With *A* amplitude, *t* time, *a* shift parameter, *b* scale parameters and *γ* asymmetry parameter.

**Time frequency representation** is used as a quantitative analysis of cortical oscillations encoding the transfer of data from one neuronal population to another.

In fact, pre ictal gamma oscillation are such random events, even statically are unpredictable. In fact, the analysis of frequency variations allows us to describe these events, beginning, end, and to recognize its variations. In this study, we used Morlet wavelet transform with oscillations wavelet number equal to 7 in order to map in the frequency domain our original signal versus pure pre ictal gamma oscillations recovered by SWT and despikifying techniques.

**Spatio-temporal representation:** We computed Morlet wavelet transform using an oscillation parameter equal to 5 (1) in order to depict a map composed of time per channels. Firstly, we emphasized pre ictal gamma oscillations band as in (18), then we normalized our map using lower frequency band, as in (19). We mapped in the time space domain our original signal versus pure pre ictal gamma oscillations obtained by SWT and despikifying in order to predict a seizure build up (time of occurrence of seizure and implicated captors and hence cortical regions).

## 3   Results

### 3.1   Simulated data

Figure1 depicts time-frequency analysis of one channel during pre ictal state of simulated data versus reconstructed oscillation by SWT and despikifying. Up box represent simulated data (mixture of spikes and oscillation), oscillations obtained by SWT and oscillations recovered by despikyfing .Lower panel are the time frequency representation of these signals. It is clear that for a mixture of oscillations and spikes there is two clear shapes pyramidal one that correspond to spikes and blobs that represent oscillations ( red circles) after filtering by SWT and despi we can see that the pyramidal shape dosnt exist any more especially for the time frequency domain of the despiky technique. This result proves that SWT is capable to detect pre ictal gamma oscillations but not in a good way (there is still a spiky shape in the recovered signal), moreover despikyfing has a better result in taking of the spiky shape and remaining pure oscillations

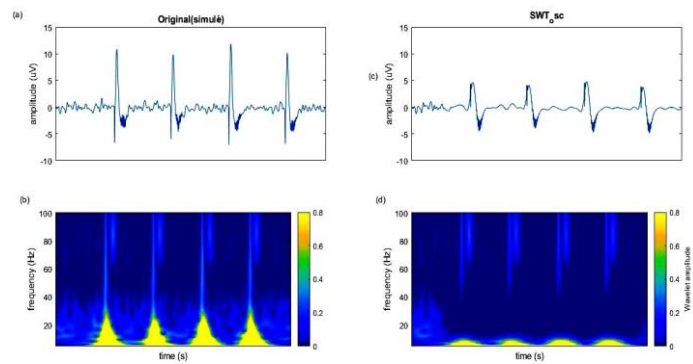

**Fig.1.** Time frequency analysis of simulated data (oscillations mixed with spikes) and recovered oscillations obtained by SWT. (a): original signal, (b): time frequency analysis of original signal, (c): filtered signal, (d): time frequency analysis of filtered signal.

Figure 2 shows spatio-temporal maps of an original simulated pre ictal data (a) and filtered data using SWT, normalized and smoothed (b). An important energy was registered in all channels for predicting seizure build up. Pre ictal gamma oscillations recovered by SWT are clearer in the entire channels, with higher significance in channel 4.

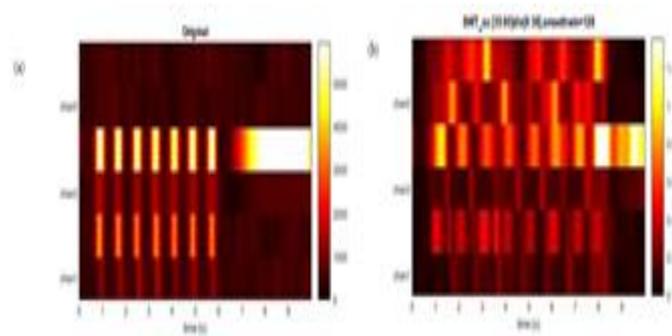

**Fig.2.** Spatio-temporal maps of simulated data: (a) Original signals, (b) recovered oscillations by SWT with normalization and smoothing.

**Despikifying.** Figure3 illustrates our original simulated data (channel 4) in the temporal domain versus our despikified filtered signal (only pre ictal gamma oscillations).

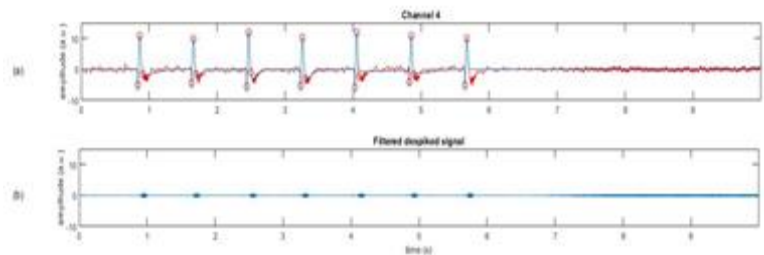

**Fig.3.** Despikifying procedure for channel 4, a) correspond to simulated original in blue and the modeling spikes in red (spikes detection shown as circles (red). (b) despiked data filtered in 65-85 Hz.

There is no clear spiky event in the despikifyied signal, only pure pre ictal gamma oscillations that exists. In figure 4, we depict time frequency map of simulated channel 4 versus the despikifyied data of channel 4. There is no more pyramidal shape in time frequency map of despikifyied signal moreover oscillatory part is much visible.

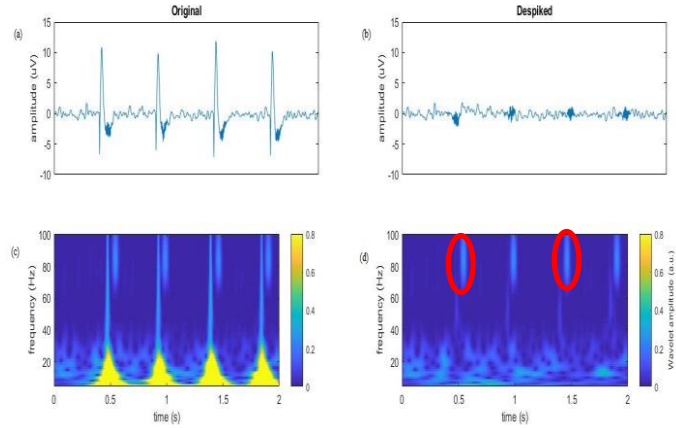

**Fig.4.** Time frequency analysis of channel 4(simulated data) : (a): original simulated signal, (b): despiked signal, (c)time frequency analysis of original signal, (d) time frequency analysis of despiked signal, no pyramidal shape, blobs that correspond to oscillations are more visible.

In figure 5 we depict spatio-temporal map of our original simulated data, versus despikifyied results. We proceed , with the same way of analyzing spatio temporal map as figure 3, original data , despikifyied data filtered in the band of frequency 65-85 Hz, , normalized by low frequency band , then finally smoothed.

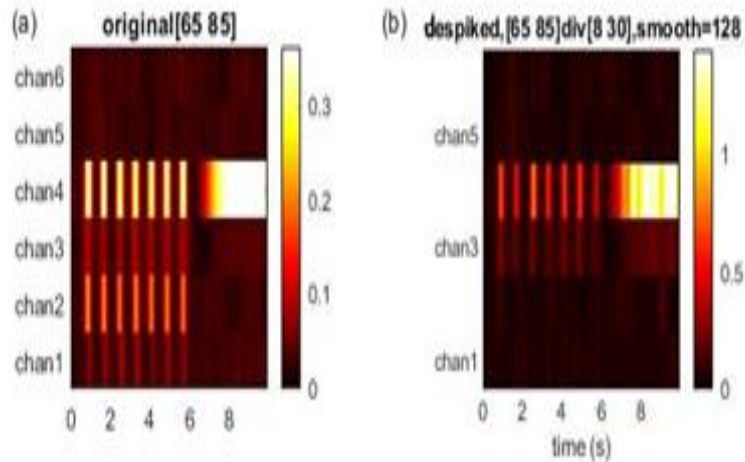

**Fig.5.** Spatio-temporal maps of simulated data: (a) originals signals filtered in 65-85 Hz band, (b) despiked signal, normalized by low frequencies 8-30 Hz, and smoothed.

### 3.2 Real data

We proceed in the same way as section simulated data to evaluate SWT versus despikifying in reconstructing of pure pre ictal gamma oscillations and predicting seizure build up, through time frequency analysis and time space mapping in order to recognize a seizure build up in time and space sensors.

We depict in figure 6 time frequency mapping of our real IEEG signal for channel Li1 versus oscillatory activities recovered by SWT and despikifying, it's confirmed that SWT still reconstruct much more spiky events with oscillatory activities , in fact in the frequency domain the pyramidal shape remains clearer in SWT, it s also to be noted that despikifying has also remained a part of spiky events and this could be justified that the spiky model couldn't reconstruct all the kind of transitory activities.

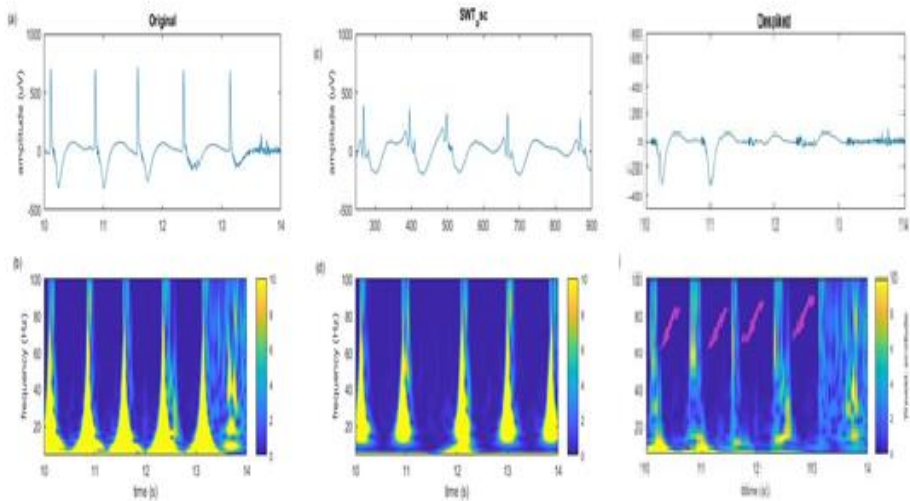

**Fig.6**. Time frequency analysis of channel li. a) real data, b) SWT results, c) despiked signal.

In Figure 7, we depict time space mapping of our real data filtered in the 65-85Hz band (a), oscillations recovered by SWT filtered normalized and smoothed (b) and despikifyied signal filtered normalized and smoothed (c).

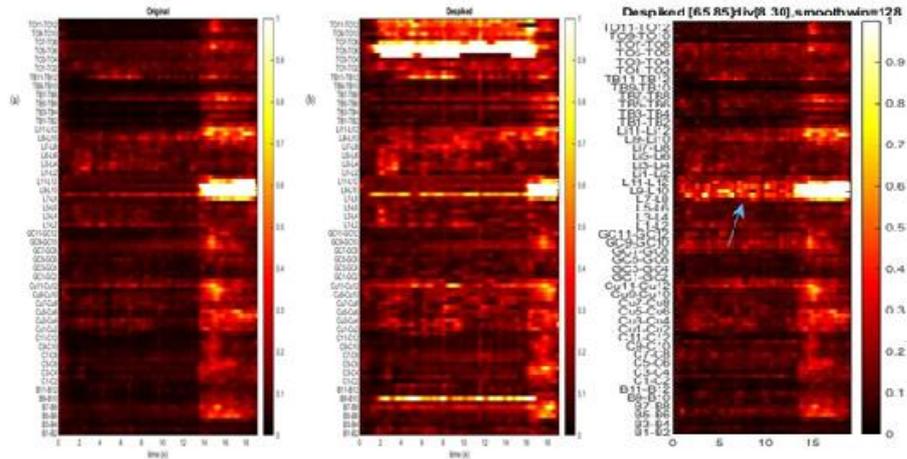

**Fig.7.** Spatio-temporal maps of: a) real signals b) SWT oscillations filtered in 65-85 Hz band normalized and smoothed, c) despikifyied oscillations filtered in 65-85 Hz band normalized and smoothed.

Seizure build up is much more clear for both recovered pure pre ictal gamma oscillations using SWT and despikifying techniques, it's about sensors Li1 to starting at 15s from our recording. Nevertheless, it's quite important to reveal that gamma oscillations reconstructed by SWT are still overlapped with spiky events which would explain the important energy seen in several sensors.

## 4   Conclusion

In this study, we evaluated two filtering techniques: SWT and Despikifying, in order to recover in first place pure pre ictal gamma oscillations none contaminated by spiky events. In second place, we used time frequency and space mapping to compare the robustness of these techniques in reconstruction of pure oscillations and in prediction of seizure builds up in time and space. In fact SWT was able to separate spikes from oscillations using a thresholding steps of masks (rectangular shape for oscillations and pyramidal one for spikes). However it's necessary to mention that for simulated and real data, there is also a remained spiky part in the recovered oscillations activities, these results were in concordance with (2). For Despikifying technique, there are no more spiky events in time course, neither in the frequency domain as pyramidal shape. It's clear that despikifying is more robust than SWT in recovering pure pre ictal gamma oscillations. For predicting seizure build up , despikifying was also more efficient in defining accurate time and location of excessive discharge and ictal state since pre ictal gamma oscillations are more pure and time space mapping  reveals a clear seizure build up. Hence Despikifying is a prominent technique in separation between spikes and oscillations, furthermore, this technique could be noteworthy in defining responsible generators of epileptogenic zone, it could also be efficient as a diagnostic aid tool for epilepsy diagnosis and for other neurological disease since it proves accurate results in separating both biomarker oscillatory and spiky events. We suggest in further work to compute networks connectivity among noninvasive techniques (EEG and MEG) for both biomarker (oscillations versus spikes) recovered by Despiking technique in order to define accurate epileptogenic zones and to plan explored regions by IEEG for surgery intervention of pharmaco resistant subjects.